\documentclass[12pt]{article}
\usepackage{amssymb,amsmath,epsfig,cite}
 \allowdisplaybreaks
\begin{document}
\title{\bf Quasi Static Evolution of Compact Objects in Modified Gravity}
\author{Z. Yousaf$^1$ \thanks{zeeshan.math@pu.edu.pk}, Kazuharu Bamba$^2$
\thanks{bamba@sss.fukushima-u.ac.jp}, M. Z. Bhatti$^1$
\thanks{mzaeem.math@pu.edu.pk} and U. Farwa$^1$ \thanks{ume.farwa514@gmail.com}\\
$^1$ Department of Mathematics, University of the Punjab,\\
Quaid-i-Azam Campus, Lahore-54590, Pakistan\\
$^2$ Division of Human Support System,\\ Faculty of Symbiotic Systems Science,
\\ Fukushima University, Fukushima 960-1296, Japan}

\date{}
\maketitle
\begin{abstract}
In this paper, the quasi static-approximation on the hydrodynamics
of compact objects is proposed in $f(R, T)$ gravity, where $R$ is the
scalar curvature and $T$ is the trace of stress-energy tensor, by exploring
the axial and reflection symmetric space time stuffed with anisotropic and dissipative
matter contents. The set of invariant-velocities is defined to comprehend the concept
of quasi static-approximation. As a consequence, the evolution of compact objects
is shown by analyzing the corresponding modified field, dynamical and scalar
equations in this approximation to evoke all the feasible outcomes. Furthermore,
the significance of kinematical quantities, modified heat-fluxes and scalar variables
are found through the proposed approximation.
\end{abstract}


\section{Introduction}

Einstein gravity theory is considered as the foundation of
cosmology and relativistic astrophysics. The observational
consequences such as $\Lambda$-cold dark matter turn out to be
stable with varied cosmological controversies other than certain
deviations like cosmic coincidence and fine-tuning
\cite{weinberg1989cosmological, peebles2003cosmological}. The
surveys of cosmic-microwave background radiations (CMBR), red-shift,
Supernovae Type Ia and very large scale-structures are the evidence
of accelerated-expansion of our cosmos
\cite{pietrobon2006integrated, giannantonio2006high, riess2007new}.
All such observations affirmed the existence of anti-gravity
entitled as dark energy (DE). The DE is a kind of energy with
enough-negative pressure and have repulsive nature of gravity whose
existence is taking for guaranteed to illustrate the observed
accelerated-expansion of our cosmos \cite{riess1998observational,
perlmutter1999measurements}. In order to modify Einstein gravity theory, various
mathematical-models have been introduced to describe the DE and
dark matter. The dark matter does not interact with
ordinary matter and is basically invisible to light. This is the
reason to call that type of matter as Dark.

The current physical cosmology is dominated by DE era
\cite{collaboration2014planck}, which illustrates the accelerated
expanding nature of universe. On behalf of its mysterious nature,
the investigation of DE is regarded as the most assertive field of
research in cosmology. The modified gravity theories (MGTs) are very
good approach to deal with that type of force which is accountable
for current accelerated speed up of universe. In last decades,
various number of MGTs are presented to comprehend the current
cosmic-epochs. The models of MGTs are proposed by modifying the
geometric section of Einstein-Hilbert action (for detailed
reviews on MGTs and DE, see, for illustration
\cite{copeland2006dynamics, padmanabhan2008dark, durrer2008dark, sotiriou2010f,
capozziello2011extended, nojiri2011unified, faraoni2010landscape,
bamba2012dark, clifton2012modified, joyce2015beyond,
bamba2015inflationary, nojiri2017modified,
yousaf2017role,1}). Nojiri and Odintsov \cite{nojiri2003modified}
presented $f(R)$ gravity and analyzed this theory is well consistent
to understand the accelerated-expansion of our cosmos. The MGTs
comprise $f(R)$ \cite{bhatti2017dynamical, bhatti2016influence,
bhatti2019tolman}, $f(G)$ \cite{nojiri2005modified,
bamba2017energy}, $f(R, T)$ \cite{harko2011f, yousaf2018existence}
(where $G,~R$ and $T$ describe Gauss Bonnet-invariant,
Ricci-scalar and the trace of stress-energy tensor, respectively)
and $f(R, T, Q)$ \cite{haghani2013further, yousaf2016stability,
yousaf2017stability, yousaf2017role, yousaf2018dynamical} theories
(here $Q=R_{\lambda\omega}T^{\lambda\omega}$, including non-minimal
coupling related to geometric and matter contents) etc.

Stellar evolution is a process which demonstrates the changes that
astrophysical object undergoes in its lifetime. The anisotropic
nature of matter contents have great relevance in the formation and
evolution of astrophysical objects. The gravitational collapse
is the most significant part of the stellar evolution. The
investigation of dynamical characteristics of self-gravitating body
is a key problem. Therefore, the study of gravitational collapse has gained wide
attention in Einstein gravity theory \cite{di1996radiating, herrera2004dynamics} as well
as in MGTs \cite{bhatti2017role, bhatti2020spherical}. Herrera and
collaborators \cite{herrera2005cylindrical, di2007nonadiabatic}
explored Some dynamical structures for cylindrical as well as
spherical collapsing fluids after the evaluation of matching
conditions. Yousaf and Bhatti \cite{yousaf2016cavity} specified the
instability constraints and the appearance of cavity in relativistic
interiors and studied the effects of dark-source terms of modified
gravity on the unstable regions. Bhatti and Yousaf
\cite{bhatti2017gravitational} also examined the effects of $f(R)$
models on the dynamical configuration of collapsing body and studied
that the collapsing process slows down because of charge and
constituents of $f(R)$ gravity.

Scalar variables (structure-scalars) have notable significance to
apprehend the physical aspects of self-gravitating stellar systems.
This idea has gained more attention of astrophysicists. In 2009,
Herrera and collaborators \cite{herrera2009structure} presented a
detailed study on relativistic set of equations for spherical
configuration controlled by scalar-variables called as structure
scalars. After this, they \cite{herrera2012cylindrically} utilized
the same idea for $(1+3)$ cylindrical formalism, and figured out
four-set of scalar variables , they also related these scalars to
the basic physical aspects of anisotropic matter contents. Yousaf
\cite{yousaf2019role} has examined the role of modified
scalars-variables in the context of $f(G, T)$ gravity and studied
the influence of such modified scalars on the evolutions of
kinematical quantities. It was also examined that these scalars are
beneficial for the review of Penrose Hawking-singularity. In 2016,
Yousaf et al. \cite{yousaf2016influence,2} studied the influence of
extended gravity on the dynamical behavior of radiating star by
evaluating modified scalars. Recently, Bhatti \emph{et
al.}\cite{bhatti2021structure} have studied the significance of
scalar-variables for the evolution of massive stars. They have also
calculated some stellar equations in the direction of $f(R)$
gravity.

The research on dynamical study of axial-symmetric anisotropic
matter contents exists in large numbers in direction of Einstein gravity theory.
However, it is a little bit inspiring to deal with such spacetime in
MGTs. Herrera \emph{et al.} \cite{herrera2013axially} studied the
dynamics of axial-symmetric and anisotropic relativistic system by
evaluating scalar-variables in static configuration. After this
attempt they \cite{herrera2014dissipative} also generalized the same
work in order to demonstrate the evolving
axial and reflection symmetric anisotropic stellar objects and
revealed nice outcomes corresponding to physical aspects by means of
these scalars. Their contributions delivered
gravitational-radiations, heat-dissipation and flow of super energy
associated with magnetic parts of the Weyl-tensor, heat-flux
vector and vorticity, respectively. Bhatti and his collaborators
\cite{bhatti2020stability} considered axial-symmetric configuration
to analyze the stability of compact bodies by imposing perturbation
scheme in the direction of $f(R, T)$ gravity. For this purpose they
investigated Newtonian as well post-Newtonian realm for particular
$f(R, T)$ model. Recently, we \cite{yousaf2021axially}
have presented the general study on axial and reflection symmetric
sources in the onset of $f(R, T)$ gravity. The relativistic equations
for the chosen system are calculated, and the generalized transport
equation is also presented to discuss the thermodynamics of the system.
They concluded that the generalized structure scalars have a significant
role in the dynamics of the system.

In this paper, we bring out the effects of extra terms associated to $f(R, T)$ theory
of gravity, on the dynamics of evolving fluid in the quasi-static regime, by
following the program outlined in \cite{yousaf2021axially}.
This paper is outlined as
follows: We provide formalism of $f(R, T)$ gravity in section \textbf {2}.
The representation of axially-symmetric
anisotropic and dissipative source, and related
kinematical variables in section \textbf {3}, we will also discuss
$f(R, T)$ scalar variables in this section. Section \textbf{4}
covers kinematics of the system, where we would like to discuss
specific-velocities. These velocity functions have special role to
define the quasi static-approximation (QSA). The next section is
devoted to the quasi static-regime to evaluate the dynamics of
our relativistic self-gravitating system. In last section, we discuss our
findings.

\section{The $f(R, T)$ Formalism}

We take into account $f(R, T)$ gravity proposed on the basis of
non minimal coupling between system's geometry and its fluids contents.
The scalar-curvature $R$ in Einstein's gravity action function is substituted
with its generic function of scalar-curvature and trace of stress-energy tensor
i.e., $R, T$. This gravity is the extension of $f(R)$ gravity,
the $f(R, T)$ gravity includes certain quantum effects
and is regarded to be even more effective than $f(R)$ gravity.
It is highlighted that such modification in the Lagrangian may be noticed
as the additional degrees of freedom. Therefore, the equation of motion
that develops from this type of Lagrangian will be different from Einstein's
gravity. In that scenario, the cosmological constant might be omitted from
the equations describing the universe's acceleratory phase. These types of
Lagrangians are extremely important for studying dark matter and dark energy
concerns (for review, please see
\cite{bamba2012dark, joyce2015beyond, houndjo2012reconstruction,
moraes2017modeling, singh2014friedmann}). The generalized
action for $f(R, T)$ gravity is expressed as \cite{harko2011f}
\begin{align}\label{a}
S_{f(R,T)}=\frac{1}{2\kappa} \int \sqrt{-g} \left[f(R,T)+
\textit{L}_m\right]d^4x,
\end{align}
where $\textit{L}_m$ is the relative Lagrangian-density of matter
contents.
The stress energy-tensor is given as
\begin{align}\nonumber
&T_{\lambda\omega}^{(m)}=-\frac{2}{\sqrt{-g}}\frac{\delta\left(\sqrt{-g}\textit{L}_m\right)}
{\delta{g^{\lambda\omega}}},
\end{align}
applying variation on Eq.\eqref{a} with respect to metric tensor
$g_{\lambda\omega}$ and we receive the following set of equations
\begin{equation}\label{b}
R_{\lambda\omega}f_{R}-\frac{1}{2}g_{\lambda\omega}f
+\left(g_{\lambda\omega}\Box -\nabla_{\lambda}
\nabla_{\omega}\right)f_{R} =\kappa
T_{\lambda\omega}-f_{T}\left(\Theta_{\lambda\omega}+T_{\lambda\omega}\right),
\end{equation}
where $g$ is the determinant of metric tensor and $\nabla_{\lambda}$ is
the operator for covariant-derivative, while
$\Box=g^{\lambda\omega}\nabla_\lambda\nabla_\omega$ identifies d'Alembert's
operator. Also,
\begin{align}\nonumber
\Theta_{\lambda\omega}=g^{\alpha\beta}\frac{\delta{T_{\alpha\beta}}}
{\delta{g^{\lambda\omega}}}=-2T_{\lambda\omega}+g_{\lambda\omega}\textit{L}_m
-2g^{\alpha\beta}\frac{\partial^2\textit{L}_m}{\partial
g^{\lambda\omega}\partial g^{\alpha\beta}},
\end{align}
by choosing  relativistic units $c=G=1$, so for $\kappa=8\pi$ and energy density $(L_m=\mu)$,
then the expression of $\Theta_{\lambda\omega}$ becomes
\begin{align}\nonumber
\Theta_{\lambda\omega}=-2T_{\lambda\omega}+\mu g_{\lambda\omega}.
\end{align}
From Eq.\eqref{b}, the field equations in $f(R,T)$ gravity are
\begin{align}\label{2}
G_{\lambda\omega}&={{T}_{\lambda\omega}}^{\textrm{eff}}
=\frac{1}{f_R}\left[(1+f_{T})T_{\lambda\omega}^{(m)}+\mu
g_{\lambda\omega}f_{T}+(\frac{f} {2}-\frac{R}{2}f_R)
g_{\lambda\omega}+\nabla_{\lambda}\nabla_{\omega}{f_R}-g_{\lambda\omega}\Box{f_R}\right],
\end{align}
where $f\equiv f(R, T)$, $R$ is the Ricci-scalar and $T$ describes
the trace of stress energy-tensor and ($f_{R}= \frac{\partial f}{\partial R}$,
$f_{T}= \frac{\partial f}{\partial T}$) and $G_{\lambda\omega}$
represents the Einstein-tensor.

\section{Axially Symmetric Geometry and Kinematical Quantities}

We consider the axial and reflection symmetric spacetime. For this system,
the generic form of the Weyl spherical-coordinates is written as
\begin{align}\label{1}
ds^2_-=-A^2(t,r,\theta)dt^{2}+B^2(t,r,\theta)(dr^{2}
+r^2d\theta^{2})+C^2(t,r,\theta)d\phi^{2} +2G(t,r,\theta) d\theta
dt,
\end{align}
where the geometric quantities such as $A, B$ are dimensionless and
at the same time $C$ and $G$ have dimension of $r$.
With the preceding mathematical form, we suppose that our axially symmetric geometry
is occupied with anisotropic and dissipative collapsing matter distribution
\begin{align}\label{3}
T^{(m)}_{\lambda\omega}=(P+\mu)U_{\lambda}U_{\omega}+P
g_{\lambda\omega}+\Pi_{\lambda\omega}+q_{\lambda}U_{\omega}+q_{\omega}U_{\lambda},
\end{align}
where $T^{(m)}_{\lambda\omega}$ is describing some usual energy components, and
four-velocity $U_{\lambda}$ is assigned by particular observer. We are dealing the
configuration, where the fluid contents are at rest position.
In our case, we have chosen the
fluid contents to be co-moving, next
\begin{align}\label{4}
U^{\lambda}=\left(\frac{1}{A},0,0,0\right),\quad
U_{\lambda}=\left(-A,0,\frac{G}{A},0\right).
\end{align}
Now, we present the unit space-like vectors in component form as follows
\begin{align}\label{5}
K_{\lambda}=\left(0,B,0,0\right),\quad
L_{\lambda}=\left(0,0,\frac{\sqrt{r^2A^2B^2+G^2}}{A},0\right),\quad
S_{\lambda}=\left(0,0,0,C\right),
\end{align}
holding the relation
\begin{align}\label{6}
&U^{\lambda}U_{\lambda}=-K_{\lambda}K^{\lambda}
=-L_{\lambda}L^{\lambda}=-S_{\lambda}S^{\lambda}=-1,\\\label{7}
&U^{\lambda}K_{\lambda}=U^{\lambda}L_{\lambda}
=U^{\lambda}S_{\lambda}=S_{\lambda}K^{\lambda}
=K^{\lambda}L_{\lambda}=S_{\lambda}L^{\lambda}=0.
\end{align}
The unitary vectors $U_{\lambda}, K_{\lambda}, L_{\lambda},
S_{\lambda}$ generate orthonormal-tetrad ($e^{a}_{\lambda}$) such as
\begin{align}\nonumber&
{e_{\lambda}}^{(0)}=U_{\lambda}, \quad
{e_{\lambda}}^{(1)}=K_{\lambda},\quad
{e_{\lambda}}^{(2)}=L_{\lambda},\quad
{e_{\lambda}}^{(3)}=S_{\lambda},
\end{align}
here $a=0, 1, 2,3$ and the representation of dual-vector tetrad is given by
\begin{align}\nonumber
&\eta_{(a)(b)}=g_{\lambda\omega}e_{(a)}^{\lambda}e_{(b)}^{\omega},
\end{align}
here $\eta_{(a)(b)}$ shows the Minkowski space-time. The
anisotropic tensor is defined with the help of scalar functions as
\cite{herrera2016axially}
\begin{align}\nonumber
\Pi_{\lambda\omega}=\frac{1}{3}(2\Pi_{I}+\Pi_{II})(K_{\lambda}K_{\omega}
-\frac{h_{\lambda\omega}}{3})+\frac{1}{3}(\Pi_{I}
+2\Pi_{II})(L_{\lambda}L_{\omega}-\frac{h_{\lambda\omega}}{3})
+\Pi_{KL}K_{(\lambda}L_{\omega)},
\end{align}
where
\begin{align}\nonumber
&\Pi_{I}=(2K^{\lambda}K^{\omega}-L^{\lambda}L^{\omega}
-S^{\lambda}S^{\omega})T_{\lambda\omega},\quad
\Pi_{II}=(2L^{\lambda}L^{\omega}-K^{\lambda}K^{\omega}
-S^{\lambda}S^{\omega})T_{\lambda\omega},\quad
\Pi_{KL}=T_{\lambda\omega}K^{\lambda}L^{\omega}.
\end{align}
This particular choice of above scalars is helpful
to evaluate the relevant equations in more easier and compact form.
Now, we introduce the heat-flux vector in form of two
scalars $q_{I}$ and  $q_{II}$ as follows
\begin{align}\label{8}
&q_{\lambda}=q_{I}K_{\lambda}+q_{II}L_{\lambda},
\end{align}
and it is observed that $U_{\lambda}q^{\lambda}=0$, so for in
coordinate-components \cite{herrera2014dissipative}
\begin{align}\label{9}
q_{\lambda}=\left(0,Bq_{I},\frac{\sqrt{r^2A^2B^2+G^2}q_{II}}{A},0\right).
\end{align}

\subsection{Kinematical Quantities}

In the study of self-gravitating system, the kinematical quantities
play significant role. Any celestial object undergoes different
phases such as distortion of shape, it can contract or expand.
Here, we would also like to express the shear-tensor ($\sigma_{\lambda\omega}$),
expansion-scalar $(\Theta)$ and the component of vorticity other
than the four-acceleration
\begin{align}\label{10}
a_{\lambda}=U^{\gamma}U_{\lambda;\gamma}=a_IK_{\lambda}+a_{II}L_{\lambda},
\end{align}
along with
\begin{align}\label{11}
a_I=\frac{A'}{AB};\quad a_{II}=\frac{A}{\sqrt{r^2A^2B^2+G^2}}\left[
\frac{A_{,\theta}}{A}-\frac{G}{A^2}\left(
\frac{\dot{A}}{A}-\frac{\dot{G}}{G}\right)\right],
\end{align}
here $\dot{A}=\frac{\partial A}{\partial t}$, $A'=\frac{\partial
A}{\partial r}$ and $A_{,\theta}=\frac{\partial A}{\partial
\theta}$.
The expansion scalar is a kinematical quantity which calculates the fractional
change of matter volume with respect to time. Whose mathematical form and expression for
our relativistic system is
\begin{align}\label{12}
\Theta=U_{;\lambda}^{\lambda}=\frac{AB^2r^2}{A^2B^2r^2+G^2}\left[
\frac{G^2}{A^2B^2r^2}\left(\frac{\dot{B}}{B}-\frac{\dot{A}}{A}
+\frac{\dot{C}}{C}+\frac{\dot{G}}{G}\right)+\frac{2\dot{B}}{B}+\frac{\dot{C}}{C}\right].
\end{align}
If $\Theta < 0, \Theta > 0$, then it represents contracting and expanding nature of matter
contents, respectively. However, $\Theta=0$ indicates the presence of vacuum cavity
inside the matter contents. Moreover, the shear tensor quantifies the distortion
in shape such that its volume remains
constant. Defined by
\begin{align}\label{13}
\sigma_{\lambda\omega}=\sigma_{(a)(b)}
e_{\lambda}^{(a)}e_{\omega}^{(b)}=U_(\lambda;\omega) +a_{(\lambda}
U_{\omega)}-\frac{h_{\lambda\omega}}{3}\Theta.
\end{align}
The shear-tensor $\sigma_{\mu\nu}$ may also be expressed in form of
scalar-functions $\sigma_I$ and $\sigma_{II}$ as
\begin{align}\label{14}
\sigma_{\lambda\omega}=\frac{1}{3}(2\sigma_{I}+\sigma_{II})
(K_{\lambda}K_{\omega}-\frac{h_{\lambda\omega}}{3})
+\frac{1}{3}(\sigma_{I}+2\sigma_{II})(L_{\lambda}L_{\omega}
-\frac{h_{\lambda\omega}}{3}),
\end{align}
where
\begin{align}\label{15}
&2\sigma_I+\sigma_{II}=\frac{3}{A}\left\{\frac{\dot{B}}{B}
-\frac{\dot{C}}{C}\right\},\\\label{16}
&\sigma_I+2\sigma_{II}=\frac{3}{r^2A^2B^2+G^2}\left[r^2AB^2
\left\{\frac{\dot{B}}{B}-\frac{\dot{C}}{C}
\right\}-\frac{G^2}{A}\left\{\frac{\dot{A}}{A}+\frac{\dot{C}}{C}
-\frac{\dot{G}}{G}\right\}\right].
\end{align}
Finally, the vorticity can be represented by
$\hat{\omega_{\lambda}}$ (vorticity-vector) or
$\Omega^{\omega\alpha}$ as follows
\begin{align}\nonumber
\hat{\omega}_\lambda=\frac{1}{2}\eta_{\lambda\omega\alpha\beta}U^{(\omega;\alpha)}U^{\beta}
=\frac{1}{2}\eta_{\lambda\omega\alpha\beta}\Omega^{\omega\alpha}U^{\beta},
\end{align}
where $\eta_{\lambda\omega\alpha\beta}$ is the Levi-Civita tensor
while the vorticity-tensor is
$$\Omega_{\lambda\omega}=U_{[\lambda;\omega]}+a_{[\lambda}U_{\omega]},$$
its single non-zero component is
\begin{align}\nonumber
\Omega_{12}=\frac{G}{2A}\left(\frac{2A'}{A}-\frac{G'}{G}\right).
\end{align}
Therefore, it can be written in form of unit-vectors as
\begin{align}\nonumber
\Omega_{\lambda\omega}=\left(L_{\lambda}K_{\omega}
-L_{\omega}K_{\lambda}\right)\Omega,\quad
\omega_{\mu}=-S_{\mu}\Omega,
\end{align}
scalar function $\Omega$ is given as
\begin{align}\label{17}
\Omega=\frac
{-G(\frac{2A'}{A}-\frac{G'}{G})}{2B\sqrt{r^2A^2B^2+G^2}}.
\end{align}
In case of central regular conditions, the above equation shows that
$G=0\Leftrightarrow\Omega=0$.

\subsection{The Weyl-tensor and Structure Scalars}

Since the magnetic portion of the Weyl-tensor does not vanish in the axially
symmetric setting. Therefore, we would like to introduce the electric
$(E_{\lambda\omega})$ as well as magnetic $(H_\lambda\omega)$ portions of the
Weyl-tensor $(C_{\lambda\omega\gamma\alpha})$.
Usually, these are defined as \cite{herrera2016axially}
\begin{align}\label{18}
E_{\lambda\omega}=C_{\lambda\beta\omega\alpha}U^{\beta} U^{\alpha},
\quad
H_{\lambda\omega}=\frac{1}{2}\eta_{\lambda\beta\epsilon\rho}
C^{~~~\epsilon\rho}_{\omega\alpha}U^{\beta}
U^{\alpha},
\end{align}
Thus, for Eq.\eqref{1}, these might be expressed as
\begin{align}\nonumber
E_{\lambda\omega}&=\frac{1}{3}\left(2\varepsilon_I
+\varepsilon_{II}\right)\left(K_{\lambda}K_{\omega}
-\frac{h_{\lambda\omega}}{3}\right)+\frac{1}{3}\left(\varepsilon_I+2\varepsilon_{II}\right)
\left(L_{\lambda}L_{\omega}-\frac{h_{\lambda\omega}}{3}\right)
+\varepsilon_{KL}\left(K_{\lambda}L_{\omega}+K_{\omega}L_{\lambda}\right),\\\nonumber
H_{\lambda\omega}&=H_1\left(K_{\lambda}S_{\omega}+K_{\omega}S_{\lambda}\right)+H_2
\left(L_{\lambda}S_{\omega}+L_{\omega}S_{\lambda}\right),
\end{align}
where
$h_{\omega}^{\lambda}=\delta^{\lambda}_{\omega}+U^{\lambda}U_{\omega}$,
and $\varepsilon_I,\varepsilon_{II},\varepsilon_{KL}$ and
$H_1,H_2$ are the components of electric and magnetics parts,
respectively. In the decomposition of the Riemann-tensor,
the Weyl-tensor plays a key role.

Structure scalars play important role to examine the physical aspects of the
fluid contents. To calculate these scalars for our problem,
let us take into account three-tensors
$X_{\lambda\omega}$, $Y_{\lambda\omega}$ and $Z_{\lambda\omega}$ through
Riemann-tensor for the evaluation of scalar-variables
\begin{align}\label{21}
X_{\lambda\omega}=\frac{1}{2}\eta^{~~~\epsilon\rho}_{\lambda\beta}
R^{*}_{\epsilon\rho\omega\alpha} U^{\alpha}U^{\beta} ,\quad
Y_{\lambda\omega}=R_{\lambda\beta\omega\alpha}U^{\beta}U^{\alpha},\quad
Z_{\lambda\omega}
=\frac{1}{2}\epsilon_{\lambda\epsilon\rho}R^{~~~\epsilon\rho}_{\alpha\omega}U^{\alpha},
\end{align}
with
$\epsilon_{\lambda\omega\rho}=\eta_{\beta\lambda\omega\rho}U^{\beta}$
and $R^{*}_{\lambda\omega\beta\alpha}
=\frac{1}{2}\eta_{\epsilon\rho\beta\alpha}R^{~~\epsilon\rho}_{\lambda\omega}.$
The explicit form of these tensors for our problem is calculated as
\begin{align}\label{22}
X_{\lambda\omega}=-E_{\lambda\omega}-\frac{\kappa}{f_R}\left((1+f_T)\frac{\Pi_{\lambda\omega}}{2}
-\mu\frac{h_{\lambda\omega}}{3}-(f-Rf_R)\frac{h_{\lambda\omega}}{6}\right)+\psi_1,
\end{align}
with the corresponding four scalar-variables,
\begin{align}\label{23}
X_T&=\frac{\kappa}{f_R}\left(\mu+\frac{1}{2}(f-Rf_R)\right)+\psi^{*}_1,\quad
\quad
X_{I}=-\varepsilon_I-\frac{\kappa}{2f_R}\left(1+f_T\right)\Pi_I,\\\label{24}
X_{II}&=-\varepsilon_{II}-\frac{\kappa}{2f_R}\left(1+f_T\right)\Pi_{II},\quad
\quad X_{KL}
=-\varepsilon_{KL}-\frac{\kappa}{2f_R}\left(1+f_T\right)\Pi_{KL}.
\end{align}
The scalar $X_T$ describes the trace part of $X_{\lambda\omega}$,
while the remaining ones are corresponding to unit space-like
vectors. Similarly,
\begin{align}\label{25}
Y_{\lambda\omega}=E_{\lambda\omega}-\frac{\kappa}{2f_R}\left[(1
+f_T)\Pi_{\lambda\omega}+\psi_{2}\right]
+\frac{\kappa}{3f_R}\left[\mu+3P+3(\mu+P)f_T+2(f-Rf_R)\right],
\end{align}
with
\begin{align}\label{26}
Y_T&=\frac{\kappa}{2f_R}\left[(\mu+3P)(1+f_T)+8{\mu
f_T}+4(f-Rf_R)+\psi_3\right],\quad
Y_{I}=\varepsilon_I-\frac{\kappa}{2f_R}(1+f_T)\Pi_I,\\\label{27}
Y_{II}&=\varepsilon_{II}-\frac{\kappa}{2f_R}(1+f_T)\Pi_{II},\quad
Y_{KL}=\varepsilon_{KL}-\frac{\kappa}{2f_R}(1+f_T)\Pi_{KL}.
\end{align}
Finally
\begin{align}\label{28}
Z_{\lambda\omega}&=H_{\lambda\omega}+\frac{\kappa}{2f_R}
(1+f_T)q^{\gamma}\epsilon_{\lambda\omega
\gamma}+\frac{\kappa}{2}\psi_4,
\end{align}
along with related scalar-variables
\begin{align}\label{29}
Z_{I}&=H_1-\frac{\kappa}{2f_R}(1+f_T)q_{II},\quad Z_{II}
=H_1+\frac{\kappa}{2f_R}(1+f_T)q_{II},\\\label{30}
Z_{III}&=H_2-\frac{\kappa}{2f_R}(1+f_T)q_{I},\quad
Z_{IV}=H_2+\frac{\kappa}{2f_R}(1+f_T)q_{I}.
\end{align}
The expressions of $\psi_i's$ are given in Appendix B.
The motivation to demonstrate such an analysis and to insight further
these structure scalars in the evolution of self-gravitating compact objects
arises from their various physical aspects.
Dissipation effects in the interior
region of stellar objects are defined by
generalized structure scalars as presented in Eqs.\eqref{29} and \eqref{30},
obtained from $Z_{\lambda\omega}$. Consequently, we can say that the
incorporation of $Z_{I, II, III, IV}$ has a direct correlation with
the magnetic effects of the Weyl-tensor and heat dissipation. Whereas
the evolution of expansion and shearing rate for axial and reflection symmetric
anisotropic and dissipative fluid is controlled by $Y_{T}$ and $Y_{I, II}$,
respectively (as expressed in Eqs.(62) and (63) in
\cite{yousaf2021axially}).
We argue that,
besides Einstein's gravity structure scalars, the generalized form
of such scalars are also significant in describing compact galactic configuration.
It is significant to note that super-massive and enormous compact galactic
structures in the universe exclusively exist in $f(R, T)$ gravity.
The specific choice of these scalar-variables is to evaluate the QSA of basic
modified scalar-equations, which are presented in Appendix A.
These sets of scalars define various physical aspects for the
evolution of self-gravitating celestial bodies.

\section{The Kinematics}

As an area of study, the kinematics is frequently referred to
describe the the geometry of motion and is sometimes considered as
subdivision of mathematics. A kinematical problem undertakes by
illustrating the geometry of systems and pointing out the initial
conditions of some known values of position of the systems. In order
to examine the large scale structure of cosmos, the kinematics is
deployed in astrophysics to exhibit the motion of celestial objects
such as stars, galaxies and collection of such objects. The
discussion of this section figures on the kinematical variables
distinguishing the motion of the medium unveiled in
\cite{demianski1986book}. We may defined the set of
invariant-velocities from the expression, containing space-like
triad $(e^{\lambda}_{(i)} ; i=1,2,3)$ given as ( for detailed study,
please see \cite{herrera2016axially}),
\begin{align}\label{31}
(\frac {D_T(\delta l)}{\delta l})_{(i,j)}=e^{\lambda}_{(i)}
e^{\omega}_{(j)}\left(\sigma_{\lambda\omega}+\frac{h_{\lambda\omega}}{3}\Theta
+\Omega_{\lambda\omega}\right).
\end{align}
From Eq.(\ref{31}), we get
\begin{align}\label{32}
&U_{(1)}=K^{\lambda}K^{\omega}\left(\sigma_{\lambda\omega}
+\Omega_{\lambda\omega}+\frac{h_{\lambda\omega}}{3}\Theta\right),\quad
U_{(2)}=L^{\lambda}L^{\omega}\left(\sigma_{\lambda\omega}+\Omega_{\lambda\omega}
+\frac{h_{\lambda\omega}}{3}\Theta\right),\\\label{33}
&U_{(3)}=S^{\lambda}S^{\omega}\left(\sigma_{\lambda\omega}+\Omega_{\lambda\omega}
+\frac{h_{\lambda\omega}}{3}\Theta\right),\quad
U_{(1,2)}=K^{\lambda}L^{\omega}\left(\sigma_{\lambda\omega}+\Omega_{\lambda\omega}
+\frac{h_{\lambda\omega}}{3}\Theta\right),\\\label{34}
&U_{(1,3)}=K^{\lambda}S^{\omega}\left(\sigma_{\lambda\omega}+\Omega_{\lambda\omega}
+\frac{h_{\lambda\omega}}{3}\Theta\right).
\end{align}
Using Eqs.(\ref{12}), (\ref{14}), (\ref{17}), we have
\begin{align}\label{35}
&U_{(1)}=\frac{1}{3}(\Theta+\sigma_I),\quad
U_{(2)}=\frac{1}{3}(\Theta+\sigma_{II}),\quad U_{(1,3)}=0,\quad
U_{(1,2)}=-\Omega,\\\label{36}
&U_{(3)}=\frac{1}{3}\left(\Theta-\sigma_{I}-\sigma_{II}\right),
\end{align}
satisfying the relation
\begin{align}\label{37}
U_{(1)}+U_{(2)}+U_{(3)}&=\Theta.
\end{align}
It is observed that the proper time-variation of $\delta l$ is
defined by these specific quantities. The geometrical as well a
physical demonstration of such specific quantities is controlled by
kinematical variables along with the unit space-like vectors as
presented in Eqs.(\ref{32})-(\ref{34}).

\section{The Quasi-static Regime}

In order to investigate self-gravitating celestial bodies, we may
take into consideration three feasible evolutionary regimes, namely:
static evolution, quasi-static evolution and dynamic
evolution. In static configuration, a coordinate structure can
always be selected in a way that all geometric as well as physical
quantities are free from time-like coordinate. In this case,
time-like hyper-surface (also orthogonal) killing vector is revealed
by spacetime. Afterwards, system undergoes complete dynamic phase,
where it is regarded to be out of equilibrium condition (either
dynamic or thermal). In between the two phases mentioned above,
there is quasi static-evolution. The system evolves slowly at every moment, so for it
may be regarded in state of equilibrium in such an evolution.

Consequently, we can say that system faces changes slowly on a
time-scale, this is very long as to that typical time in which the
system responds to small perturbed configuration of hydro-static
state of equilibrium. Thus , we can say that in this phase our
system is convenient to hydro-static state of equilibrium, and such
system may be evaluated in sequence of equilibrium-models. Now, we
would like to describe the conditions of the QSA inform of
kinematical quantities, specific-velocities and $f(R, T)$ corrections
discussed in above sections. These conditions are entailed due to
the fact that hydro-static time of system under consideration must
be much larger than any characteristic time-scale of that system.
Therefore
\begin{itemize}
\item All quantities having order O$(\epsilon^2)$ and higher will be
neglected, where $\epsilon<<1.$
\item The specific-velocities such as $U_{(1),(2),(3)}$ and
$U_{(1,2)}$ defined in  Eqs.(\ref{32})-(\ref{34}) are smaller
quantities, therefore have order O$(\epsilon).$
\item It follows from Eqs.(\ref{35}), (\ref{36}) that the scalars
$\Omega,\Theta,\sigma_{I,II}$ have order O$(\epsilon),$ also
Eqs.(\ref{12}), (\ref{15})-(\ref{17}) indicate that $G, \dot{A}, \dot{B}$
and $\dot{C}$ are of  O$(\epsilon).$
\item It is also observed from Eqs. (\ref{12}), (\ref{15}), (\ref{16}) that
$\tilde{\sigma}\equiv\sigma_{I}=\sigma_{II}$, having the same order
i.e, O$(\epsilon)$ and
\begin{align}\label{38}
\Theta-2\tilde{\sigma}=\frac{3}{A}\left(\frac{\dot{C}}{C}\right),\quad
\tilde{\sigma}+\Theta=\frac{3}{A}\left(\frac{\dot{B}}{B}\right)
\end{align}
\item The dark sources terms such that $f_R\equiv \tilde{f}_R, f_T\equiv
\tilde{f}_T, q^{\textrm{eff}}_I\equiv \tilde{q_{I}}^{\textrm{eff}},
q^{\textrm{eff}}_{II}\equiv \tilde{q_{II}}^{\textrm{eff}},
\mu^{\textrm{eff}}\equiv \tilde{\mu}^{\textrm{eff}}$ and other
effective fluid components in the QSA accordingly.
\end{itemize}
Moreover, it is also suppose that the relaxation-time in the evolution of modified
transport-equation must be neglected. In fact,
relaxation-time is the time required by the system to come back instinctively
in its steady state, after it has been abruptly took away from it.
However, it deduces from the nature of the QSA that all the processes
evolve on larger time-scale than the time taken for
transient-phenomena, inferring that we are expecting the modified
heat-fluxes to characterize a constant heat flow along with the
effect of $f(R, T)$ corrections. Therefore, the relaxation-time $\tau$ is
neglected in both components of modified transport-equation
(Eqs.(49) and (50) in \cite{yousaf2021axially}) for
our relativistic-system then the
results are obtained as
\begin{align}\label{39}
&{\tilde{q_{I}}}^{\textrm{eff}}=-\frac{\kappa}{B}\left(BTa_I+T'\right),\\\label{40}
&{\tilde{q_{II}}}^{\textrm{eff}}=\frac{\kappa}{A}\left(ATa_{II}
+\frac{G\dot{T}+A^2T_{,\theta}}{rAB}\right).
\end{align}
Since $\dot{T}$ has order $O(\epsilon)$, and using
thermal-equilibrium conditions \cite{tolman1930weight}. Therefore,
from above equations, we receive the following expressions in the quasi
static-regime
\begin{align}\label{41}
(TA)'=\frac{1}{rB\tilde{f_R}}\chi^{approx}_1, \quad
(TA)_{,\theta}=\frac{1}{rAB\tilde{f_R}}\chi^{approx}_2,
\end{align}
where ``approx" is used to illustrate the QSA on corresponding
quantities. One can be easily evaluated the quasi-static approximated values
of these quantities (presented in Appendix in \cite{yousaf2021axially},
by using the above defined QSA. Moreover, the
scalar components of Eq.(\ref{11}) are turned out to be of order
$O(\epsilon)$ after imposing such an approximation, in order that
\begin{align}\label{42}
a_{I}=\frac{1}{B}\left(\frac{A'}{A}\right),\quad a_{II}
=\frac{1}{rB}\left(\frac{A_{,\theta}}{A}\right).
\end{align}

\subsection{QSA on Modified Field Equations:}

Here, we would like to evaluate the MFEs in the QSA.
By implementing the proposed conditions,
as defined earlier in this section, we acquire the MFEs (Eqs.(14)-(20) in
\cite{yousaf2021axially}) as follows
\begin{align}\nonumber
G_{00}&=\frac{\kappa
A^2}{\tilde{f_R}}\left[\mu-\frac{1}{2}(\tilde{f}-\tilde{R}\tilde{f_R})\right]
+\frac{\kappa}{\tilde{f_R}}\left[\frac{1}{r^2B^2}\tilde{f}_{R,\theta\theta}
+\tilde{f}'_R\left\{\frac{A^2}{rB^2}+\frac{A^2}{B^2}\frac{C'}{C}\right\}\right.
\\\label{43}&\left.+\tilde{f}_{R,\theta}\frac{A^2}{r^2B^2}\frac{C_{\theta}}{C}
\right],\\\label{44}
G_{01}&=\frac{\kappa}{\tilde{f_R}}\left[-AB(1+\tilde{f_T})q_{I}\right],\\\nonumber
G_{02}&=\frac{-ABr\kappa}{\tilde{f}_R}\left[\frac{\mu
G}{ABr}+(1+\tilde{f}_T)q_{II}\right] +\frac{\kappa
G}{2\tilde{f_R}}(\tilde{f}-\tilde{R}\tilde{f_R})-\frac{\kappa
G}{2r^2B^2\tilde{f}_R} \left[\right.\tilde{f}_{R,\theta\theta}
\\\label{45}&\left.+\frac{C_{,\theta}}{C}\tilde{f}_{R,\theta\theta}\right],
\\\label{46}
G_{12}&=\frac{\kappa}{\tilde{f}_R}\left[(1+\tilde{f}_T)(B^2r\Pi_{KL})
+\frac{BG}{A}q_I+\tilde{f'}_{R,\theta}-\frac{B_{\theta}}{B}
\tilde{f'}_{R}-\frac{(Br)'}{Br}\right],\\\nonumber
G_{11}&=\frac{\kappa
B^2}{f_R}\left[(1+\tilde{f}_T)(P+\frac{\Pi_{I}}{3})
+\mu\tilde{f}_T+\frac{1}{2}(\tilde{f}-\tilde{R}\tilde{f_R})
+\frac{1}{r^2B^2}\left(\right.\right.\tilde{f}_{R,\theta\theta}
+(\frac{A_{\theta}}{A}-\frac{B_{\theta}}{B}
\\\label{47}&+\frac{C_{\theta}}{C}\left.\left.)\tilde{f}_{R,\theta}\right)\right],\\\nonumber
G_{22}&=\frac{\kappa
r^2B^2}{\tilde{f}_R}\left[(1+\tilde{f}_T)(P+\frac{\Pi_{II}}{3})+\mu\tilde{f}_{T}
+\frac{1}{2}(\tilde{f}-\tilde{R}\tilde{f_R})-\frac{1}{r^2B^2}\left(\tilde{f}_{R,\theta\theta}
+(\right.\right.\frac{A_{\theta}}{A}-\frac{B_{\theta}}{B}
\\\label{48}&\left.\left.+\frac{C_{\theta}}{C})\tilde{f}_{R,\theta}\right)\right],\\\nonumber
G_{33}&=\frac{\kappa
C^2}{\tilde{f}_R}\left[(1+\tilde{f}_T)\left(P-\frac{1}{3}(\Pi_I+\Pi_{II})\right)
+\mu\tilde{f}_{T}+\frac{1}{2}(\tilde{f}-\tilde{R}\tilde{f_R})
-\frac{1}{r^2B^2}\left(\tilde{f}_{R,\theta\theta}\right.\right.
\\\label{49}&\left.\left.+\frac{A_{\theta}}{A}\tilde{f}_{R,\theta}\right)\right],
\end{align}
here $f_{R,\theta}=\frac{\partial f_R}{\partial \theta}$ and
$B_{,\theta}=\frac{\partial B}{\partial\theta}$.

\subsection{QSA on Hydro-dynamics:}

The dynamical equations describe the change in the parameters of physical
system with respect to time. These equations are related to the study
of motion of celestial objects which is supported by stress-energy tensor.
As a result of gravitational collapse,
static celestial objects become un-stable. To deal such situation, the
gravitational field equations are helpful to provide the dynamical equations.
Therefore, we want to execute the evolution of dynamical equations in the quasi
static constraints. Thus, the quasi static-configuration of dynamical
equation (Eq.(54) in \cite{yousaf2021axially}) is given as
\begin{align}\nonumber&
\frac{1}{\tilde{f_R}}\left[(1+\tilde{f_T})\left\{\frac{\dot{\mu}}{A}+\Theta(\mu+P)
+\frac{1}{9}\left(\Pi_I(2\sigma_I
+\sigma_{II})+\Pi_{II}(\sigma_I+2\sigma_{II})\right)+\frac{q'_I}{B}
+\frac{1}{Br}(\right.\right.q_{,\theta}
\\\label{50}&+\frac{G}{A^2}\dot{q_{II}}\left.\left.)
+2(q_Ia_I+q_{II}a_{II})
+\frac{q_I}{B}(\frac{C'}{C}+\frac{(Br)'}{Br})++\frac{q_{II}}{Br}(\frac{B_{,\theta}}{B}
+\frac{C_{,\theta}}{C})\right\}\right]=\frac{1}{\tilde{f_R}}\chi^{approx}_7.
\end{align}
From the modified Euler-lagrange equation (Eq.(55) in \cite{yousaf2021axially}), the following two
equations are attained in the QSA
\begin{align}\nonumber&
\frac{1}{\tilde{f_R}}\left[(1+\tilde{f_T})\left\{\frac{1}{B}\left(P+\frac{\Pi_I}{3}\right)'+
\frac{1}{Br}\left(\Pi_{KL,\theta}+\frac{G}{A^2}\dot{\Pi_{KL}}\right)+(\mu+P+\frac{\Pi_I}{3})a_I
+a_{II}\Pi_{KL}\right.\right.
\\\nonumber&\left.\left.+\frac{\Pi_I}{3B}\left(\frac{2C'}{C}+\frac{(Br)'}{Br}\right)
+\frac{\Pi_{II}}{3B}\left(\frac{C'}{C}-\frac{(Br)'}{Br}\right)
+\frac{\Pi_{KL}}{Br}\left(\frac{2B_{,\theta}}{B}+\frac{C_{,\theta}}{C}\right)
\frac{\dot{q_I}}{A}\right\}\right]
\\\label{51}&=\left(-\frac{1}{2\tilde{f_R}}(\tilde{f}-\tilde{R}\tilde{f_R})\right)'+
\frac{1}{B\tilde{f_R}}\chi^{approx}_8,
\end{align}
and
\begin{align}\nonumber&
\frac{1}{\tilde{f_R}}\left[(1+\tilde{f_T})\left\{\frac{1}{Br}\left((P
+\frac{\Pi_I}{3})_{,\theta}+\frac{G}{A^2}(\dot{P}+\frac{\dot{\Pi_{II}}}{3})\right)
+\frac{\Pi'_{KL}}{B}+(\mu+P+\frac{\Pi_{II}}{3})a_{II}
+a_{I}\Pi_{KL}\right.\right.
\\\nonumber&\left.\left.+\frac{\Pi_I}{3Br}\left(-\frac{B_{,\theta}}{B}+\frac{C_{,\theta}}{C}\right)
+\frac{\Pi_{II}}{3Br}\left(\frac{B_{,\theta}}{B}+\frac{2C_{,\theta}}{C}\right)
+\frac{\Pi_{KL}}{B}\left(\frac{C'}{C}+\frac{(Br)'}{Br}\right)
\frac{\dot{q_{II}}}{A}\right\}\right]
\\\label{52}&=\frac{1}{\tilde{f_R}}\left[\frac{1}{Br}(\tilde{R}\tilde{f_R}-\tilde{f})
-\chi^{approx}_9\right].
\end{align}

\subsection{QSA on Modified Scalar Equations:}

In our regime, it is followed from Eq.(\ref{17}) that the time
derivative of vorticity-scalar i.e, $\dot{\Omega}$ have order
$O(\epsilon^2)$. The evolution of generalized Ricci-Identities
that are Eqs.(66) and (67) in \cite{yousaf2021axially},
in quasi static constraints yields respectively,
\begin{align}\label{53}
&\frac{2}{3B}\Theta'-\frac{\Omega}{Br}\left(
\frac{2A_{,\theta}}{A}+\frac{C_{,\theta}}{C}\right)
-\frac{\Omega_{,\theta}}{Br}-\frac{\tilde{\sigma}'}{3B}
-\frac{\tilde{\sigma}C'}{BC}=\kappa
\tilde{q_{I}}^{\textrm{eff}},\\\label{54}
&\frac{2}{3Br}\Theta_{,\theta}+\frac{\Omega'}{B}-\frac{\Omega}{B}\left(
\frac{2A'}{A}+\frac{C'}{C}\right)-\frac{\tilde{\sigma}_{,\theta}}{3Br}
-\frac{\tilde{\sigma}C_{,\theta}}{rBC}=\kappa
\tilde{q_{II}}^{\textrm{eff}}.
\end{align}
It depicts that dissipative fluxes have also order $O(\epsilon)$. So
for, we summarize all the outcomes deduced from the QSA as
\begin{itemize}
\item Order of $\dot{\Omega}, \dot{G}$ is $O(\epsilon^2)$.
\item $\tilde{\sigma}, \Omega, \Theta, G, \dot{A},
\dot{B}, \dot{C}, \dot{a_I}, \dot{a_{II}}$
are of order $O(\epsilon)$.
\item $q_I, q_{II}, \tilde{q_{I}}^{\textrm{eff}},
\tilde{q_{II}}^{\textrm{eff}}, \tilde{f_R},
\tilde{f_T}$ all are of order $O(\epsilon)$.
\end{itemize}
Since the hydro-static equilibrium state can be detained at any
time, therefore the corresponding equations containing $\sigma_{22}, \sigma_{33}$
components of $\sigma_{\lambda\omega}$ hold \cite{herrera2014dissipative}.
It is obtain from above mentioned equations, respectively
\begin{align}\label{55}
&\dot{\Pi}_{KL}\approx O(\epsilon);\quad \dot{q}_{I}\approx
O(\epsilon^2); \quad \ddot{C}\approx O(\epsilon^2); \quad
\ddot{B}\approx O(\epsilon^2),\\\label{56} &\dot{\Pi}_{II}\approx
O(\epsilon); \quad \dot{P}\approx O(\epsilon); \quad
\dot{q}_{II}\approx O(\epsilon^2).
\end{align}
It has been imposed the fact that $P, \Pi_{I}, \Pi_{II}$ include
terms with $\ddot{C}$ and $\ddot{B}$ other than the terms involving
some spatial-coordinate derivatives of corresponding line-element.
Now, it is followed immediately from Eqs.(\ref{38})
\begin{align}\label{57}
&\dot{\Theta}\approx O(\epsilon^2); \quad \dot{\tilde{\sigma}}\approx O(\epsilon^2).
\end{align}
By using  Eqs.(\ref{35}), then the Eq.(\ref{53}) turned out to be
\begin{align}\label{58}
&2U'=\frac{\Omega}{r}[\ln(\Omega CA^2)]_{,\theta}+\tilde{\sigma}[\ln(\tilde{\sigma}C)]'
+\kappa B{\tilde{q_{I}}}^{\textrm{eff}},
\end{align}
where $U\equiv U_{1}\equiv U_{2}$, after integration we attain
\begin{align}\label{59}
&U=U_{\Sigma}-\frac{1}{2}\int^{r_\Sigma}_{r}\left\{\frac{\Omega}{r}[\ln(\Omega CA^2)]_{,\theta}
+\tilde{\sigma}[\ln(\tilde{\sigma}C)]'+\kappa B{\tilde{q_{I}}}^{\textrm{eff}}\right\}dr,
\end{align}
it can also be expressed as
\begin{align}\label{60}
&U_{3}=U_{(3)\Sigma}-\frac{1}{2}\int^{r_\Sigma}_{r}
\left\{\frac{\Omega}{r}[\ln(\Omega CA^2)]_{,\theta}
+\tilde{\sigma}\left[\ln(\frac{C}{\tilde{\sigma}})\right]'
+\kappa B{\tilde{q_{I}}}^{\textrm{eff}}\right\}dr,
\end{align}
here, the equation $r=r_\Sigma$ described the surface boundary of
the source and we have also used the fact that
$U_{3}=U-\tilde{\sigma}$. In similar fashion, Eq.(\ref{54}) may be
written as follows
\begin{align}\label{61}
&2U_{,\theta}=-\Omega r[\ln(\frac{CA^2}{\Omega})]'
+\tilde{\sigma}[\ln(\tilde{\sigma}C)]_{,\theta}
+\kappa Br{\tilde{q_{II}}}^{\textrm{eff}},
\end{align}
generating
\begin{align}\label{62}
U=U_\Sigma-\frac{1}{2}\int^{\theta_\Sigma}_{\theta}\left\{-\Omega
r[\ln(\frac{CA^2}{\Omega})]'
+\tilde{\sigma}[\ln(\tilde{\sigma}C)]_{,\theta} +\kappa
Br{\tilde{q_{II}}}^{\textrm{eff}} \right\}d\theta,
\end{align}
or
\begin{align}\label{63}
U_{(3)}=U_{(3)\Sigma}-\frac{1}{2}\int^{\theta_\Sigma}_{\theta}
\left\{-\Omega r[\ln(\frac{CA^2}{\Omega})]'
+\tilde{\sigma}\left[\ln(\frac{C}{\tilde{\sigma}})\right]_{,\theta}
+\kappa Br{\tilde{q_{II}}}^{\textrm{eff}}
\right\}d\theta,
\end{align}
in this case, the boundary surface is given by the equation
$\theta=\theta_\Sigma$. Now, we will focus on the physical
description of the Eqs.\eqref{59}, \eqref{60}, \eqref{62}, \eqref{63}.
Let us investigate the order of the magnetic part of the
Weyl-tensor. From Eqs.(68) and (69) in \cite{yousaf2021axially},
and the final outcomes in that case
are same as presented in \cite{herrera2016axially}
\begin{align}\label{64}
&H_1=-a_{I}\Omega-\frac{1}{2Br}\left(\frac{\tilde{\sigma}
C_{,\theta}}{C}+\tilde{\sigma}_{,\theta}\right)
+\frac{1}{2B}\left(\frac{\Omega C'}{C}-\Omega'\right)\\\label{65}
&H_2=-a_{II}\Omega+\frac{1}{2Br}\left(\frac{\Omega
C_{,\theta}}{C}-\Omega_{,\theta}\right)
+\frac{1}{2B}\left(\frac{\tilde{\sigma}
C'}{C}+\tilde{\sigma}'\right),
\end{align}
inferring that the $H_1$ and $H_2$ are of order 0$(\epsilon)$. It is
worth observing that $G=0=\Omega$ i.e, in vorticity free case, it
implies from Eqs.(\ref{64}), ({65}) that
\begin{align}\label{66}
H_1=-\frac{(\tilde{\sigma C}_{,\theta})}{2rBC}, \quad
\quad H_2=-\frac{(\tilde{\sigma C})'}{2BC}.
\end{align}
Then from Eqs.(\ref{53}), ({54}), using Eqs.(\ref{38}) and with the
assumption $\Omega=0$, yield respectively
\begin{align}\label{67}
2\left(\frac{\dot{B}}{AB}\right)'-\frac{(\tilde{\sigma}C)'}{C}=\kappa
B\tilde{q_{I}}^{\textrm{eff}},\\\label{68}
2\left(\frac{\dot{B}}{AB}\right)_{,\theta}-\frac{(\tilde{\sigma}C)_{,\theta}}{C}=\kappa
Br\tilde{q_{II}}^{\textrm{eff}}.
\end{align}
Now, combining both above equations with Eq.(\ref{66}), we attain
the following relations, respectively
\begin{align}\label{69}
&H_1+\frac{1}{rB}\left(\frac{\dot{B}}{AB}\right)_{,\theta}
=\frac{\kappa}{2}{\tilde{q}_{II}}^{\textrm{eff}},\\\label{70}
&H_2-\frac{1}{B}\left(\frac{\dot{B}}{AB}\right)'
=-\frac{\kappa}{2}{\tilde{q}_{I}}^{\textrm{eff}}.
\end{align}
Thus from Eq.\eqref{66} it is followed that disappearance of shear
$(\Omega=0)$ is the necessary as well as sufficient condition for
the matter to be purely-electric in the QSA. The expressions of modified heat-fluxes
and the other fluid contents are given in Appendix B. However, the modified
heat-fluxes have also played an effective role in that case as well.
The quasi static-configuration of modified scalar equations is explicitly
written in Appendix A.

\section{Conclusion}

In current research, the compact objects are the most significant
class of astrophysical objects. These objects are extremely dense
i.e, smaller in size and higher in mass. The study of such objects
have gained the main attention of astrophysicists. In this paper,
we have investigated the quasi static-evolution of
axially and reflection symmetric fluids using the frame-work constructed in
\cite{yousaf2021axially}.
We have proposed axial and reflection
symmetric system, stuffed with anisotropic as well as
dissipative fluid contents. So for we have chosen the most generic
representation of stress energy-tensor as given in Eq.(\ref{3}). We developed MFEs to
study the motion of relativistic source. To analyze the basic aspects
of fluid contents, we discussed the
kinematical quantities including $a_{\lambda}$ (four-acceleration),
$\Theta$ (expansion-scalar), $\Omega$ (vorticity-scalar) and
$\sigma_{\lambda\omega}$ (shear-tensor). Moreover,
we have discussed the magnetic and electric parts
of the Weyl-tensor. Five scalars $H_1, H_2$ and $\varepsilon_I,
\varepsilon_{II}, \varepsilon_{KL}$ illuminating the magnetic and
electric-parts, respectively, are also established.

For the evolution of compact objects, three feasible regimes may be
considered, namely: static, quasi-static and dynamic regimes. The
QSA is the sensible approach to discuss the hydro-dynamics of
self-gravitating compact objects. In this evolutionary phase, the
system faces changes sufficiently slow, so for it can be regarded
in equilibrium state. For this scenario
\begin{itemize}
\item Firstly, the
set of invariant-velocities are defined for the comprehension of kinematics as
well as for the concept of the QSA.
These scalar functions basically hold the relation
\begin{align}\nonumber
U_{(1)}+U_{(2)}+U_{(3)}&=\Theta.
\end{align}
It can be seen from Eqs.\eqref{32}-\eqref{34} that
geometrical and physical demonstration of such specific-velocities is governed by
kinematical-variables with the unit space-like vectors
$K^{\lambda}, L^{\lambda}$ and $S^{\lambda}$.
Herrera \emph{et al.}
\cite{herrera2016axially} used this approach for the study of
axial and reflection symmetric self-gravitating anisotropic
and dissipative source in Einstein gravity theory.
\item The set of seven MFEs for our $(1+3)$ formalism are calculated
by using Eq.\eqref{2}, then quasi static
constraints defined in Sec. $\textbf{4}$ are imposed to evaluate
the proposed approximation.
The quasi static-configuration of these MFEs are presented in
Eqs.\eqref{43}-\eqref{49}. The continuity as well
as generalized Euler-equation are also evaluated in this configuration.
Two equations Eqs.\eqref{51} and \eqref{52}
are obtained from generalized Euler-equation in the QSA,
containing the extra-curvature terms due
to the effects of $f(R, T)$ gravity as compared to the results presented
in Einstein's gravity theory \cite{herrera2016axially}.
\item The modified heat-fluxes are also executed to examine
the thermodynamic aspects of self-gravitating evolving fluid through
the proposed approximation. The significant role played
by such kinematical quantities as well as modified heat-fluxes is
clearly revealed through the Eqs.\eqref{59}, \eqref{60}, \eqref{62} and
\eqref{63}. The above mentioned constituents lead to the pattern of
various structures. It is to be noted that in the scenario of $f(R, T)$ gravity,
the modified heat-fluxes and extra curvature terms are emerged as shown in
Eqs.\eqref{39}, \eqref{40}, \eqref{41}, however in Einstein's
gravity theory \cite{herrera2016axially},
the Eqs.\eqref{39} and \eqref{40} were governed by usual heat-flux
components while the result of
Eq. (\ref{39}) was depicted as
\begin{align}\nonumber
(TA)'=0, \quad
(TA)_{,\theta}=0
\end{align}
\item Most important, the magnetic-part of the
Weyl-tensor is not vanished in the QSA. However, it is followed
immediately from Eqs.\eqref{64} and \eqref{65} that temporal derivatives
of $H_1$ and $H_2$ is at least of order $O(\epsilon^2)$ i.e,
$\dot{H_2}\approx\dot{H_1}\approx O(\epsilon^2)$, consequently not
to be worth considering and are neglected in the QSA accordingly.
This suggests that if magnetic-part of the Weyl-tensor disappears at
any moment, afterwards the same situation will appear at any time.
The transformation of energy is carried by the modified heat-fluxes,
these effective constituents express the corrections of $f(R, T)$
gravity.
\item It is noticed that, when fluid contents are considered non
dissipative, shear-free and
irrotational, then the sign of $U$ and $U_{(3)}$ is
identical to the sign of $U_\Sigma$ and $U_{(3)\Sigma}$, however the
influence of $f(R, T)$ constituents illustrate a steady heat-flow.
Even so in the emergence of any of the above factors (Shear,
vorticity, modified heat-flux), the system may lead to a position
where ever velocity swaps of sign in matter distribution according
to its sign on surface boundary, along with the effective components
of heat-flux. These effective constituents play a productive role to
understand the behavior of such velocity functions for the quasi
static-evolution of massive objects. Consequently, it can be happened
that outer regions move in a direction opposite to that of inner
ones which propagate in mono direction.
\end{itemize}
The role of generalized scalar-variables is analyzed in the
dynamics of self-gravitating compact objects. It is also observed
that one of $f(R, T)$ scalar variables, $X_T$ demonstrates the energy
density of the matter composition along with the additional curvature
$f(R, T)$ constituents, however its irregularity in terms of local
isotropy is well-described via the remaining ones, which are $X_I$,
$X_{II}$, and $X_{KL}$.
Finally, the quasi static-evolution of $f(R, T)$ scalar equations
is presented in Appendix A, which depicts the salient physical aspects
of scalar-variables along with effective fluid contents.

The considerable number of stars end up spending a lot of their active lifetimes in
this state of equilibrium, fusing hydrogen into helium, yet it is the steady
transition of elements through the fusion mechanism that allows their setup to
change in any significant way. To investigate self-gravitating stellar objects, we may
take into consideration three feasible evolutionary regimes, in between the static and
dynamic evolution, we may have the quasi-static regime. This is the regime where
the system is considered to evolve gradually slow so that it can be assumed to be in state
of equilibrium at every moment. This indicates that the system evolves slowly on a time-scale
larger than the typical one for that fluid responds to a small perturbative configuration
of hydro-static equilibrium. The hydrostatic time scale is the term applied to this
type of time scale \cite{hansen2012stellar}. Consequently, one can say that the
stellar objects are in hydro-static equilibrium in this phase. The quasi-static
approach is very effective, for several stages of the life of the celestial bodies
\cite{kippenhahn1990stellar},
because the hydro-static time frame is of the order $10^{-4}$ seconds for
a neutron-star and $4.5$ seconds for a white-dwarf and more important
$27$ minutes for the sun. As
hydro-static equilibrium is the state in which a gaseous objects's internal
pressure exactly balances its gravitational pressure, such as celestial objects.
In order to discuss the quasi-static evolution of compact objects, we
define the QSA in form of kinematical variables, invariant velocities and $f(R, T)$
modifications, as presented in section \textbf{5}. It is worth referring that $\gamma=0$
bring out the outcomes of Einstein gravity theory as presented in
\cite{herrera2016axially}.

\vspace{0.3cm}

%%%%%%%%%%%%%%%%%%%%%%%%
%%%  Acknowledgments
%%%%%%%%%%%%%%%%%%%%%%%%
\section*{Acknowledgments}

The work of ZY, MZB and UF was supported financially by University of the Punjab research project, 2021. The work of KB
was supported in part by the JSPS KAKENHI Grant Number JP21K03547.

\vspace{0.3cm}

\vspace{0.25cm}

\section*{Funding and/or Conflicts of interests/Competing interests}

The author declares that he has no known competing financial interests or personal relationships that could have appeared to influence the work reported in this paper. Funding information is not applicable.

\section*{Data availability statement}

All data generated or analysed during this study are included in
this published article (and its supplementary information files).

\section{Appendix A}

Here, we are interested to discuss the evolution of $f(R, T)$
scalar-equations (70)-(78) in \cite{yousaf2021axially} in the QSA as follows
\begin{align}\nonumber
&\frac{1}{3A}\left(\varepsilon_I+\frac{\kappa}{2\tilde{f_R}}(1+\tilde{f_T})(\Pi_I+\mu)\right)^{\dot{}}
+\frac{1}{3}(\Theta\varepsilon_I+\tilde{\sigma}\varepsilon_{II})
-\Omega\left(\varepsilon_{KL}+\frac{\kappa}{2\tilde{f_R}}(1+\tilde{f_T})\Pi_{KL}\right)
-\frac{1}{Br}
\\\nonumber&\times\left(H_{1,\theta}+H_1\frac{C_{,\theta}}{C}\right)
-\frac{H_2}{B}\left(\frac{C'}{C}-\frac{(Br)'}{Br}\right)
=2a_{II}H_1-\frac{\kappa}{6}(\Theta+\tilde{\sigma})\left(\tilde{\mu}^{\textrm{eff}}
+(\tilde{P}+\frac{\tilde{\Pi_I}}{3})^{\textrm{eff}}\right)
\\\label{71}&-\kappa a_{I}\tilde{q_{I}}^{\textrm{eff}}-\frac{\kappa}{2B}
\left[\tilde{q_{I}}^{\textrm{eff}}
+\frac{B_{,\theta}}{B}\tilde{q_{II}}^{\textrm{eff}}\right],\\\nonumber
&\frac{1}{A}\left(\varepsilon_{KL}+\frac{\kappa}{2\tilde{f_R}}(1+\tilde{f_T})\Pi_{KL}\right)^{\dot{}}
+\frac{\Omega}{6}\left[\varepsilon_{I}-\varepsilon_{II}+\frac{\kappa}{2\tilde{f_R}}(1+\tilde{f_T})(\Pi_{I}-\Pi_{II})
\right]-(a_{II}H_2
\\\nonumber&-a_{I}H_1)-\left(\varepsilon_{KL}+\frac{\kappa}{2\tilde{f_R}}(1+\tilde{f_T})\Pi_{KL}\right)
(\tilde{\sigma}-\Theta)-\frac{1}{2B}\left[H_1\left(\frac{(Br)'}{Br}-\frac{2C'}{C}\right)-H'_1\right]
\\\nonumber&-\frac{1}{2Br}\left[H_{2,\theta}-H_2\left(\frac{B_{,\theta}}{B}\right.\right.
\left.\left.-\frac{2C_{,\theta}}{C}\right)\right]
=-\frac{2\kappa}{6}(2\tilde{\sigma}-\Theta)\tilde{\Pi_{KL}}^{\textrm{eff}}-\frac{\kappa}{2}
\left(a_{II}\tilde{q_{I}}^{\textrm{eff}}+a_{I}\tilde{q_{II}}^{\textrm{eff}}\right)
\\\label{72}&-\frac{\kappa}{4B\tilde{f_R}}(1+\tilde{f_T})\left(q'_{II}-q_{II}\frac{(Br)'}{Br}\right)
-\frac{\kappa}{4rB\tilde{f_R}}(1+\tilde{f_T})\left(q_{I,\theta}
-q_{I,\theta}\frac{B_{,\theta}}{B}\right),\\\nonumber
&\frac{1}{3A}\left[\varepsilon_{II}+\frac{\kappa}{2\tilde{f_R}}(1+\tilde{f_T})(\Pi_{II}+\mu)\right]^{\dot{}}
+\frac{1}{3}(\Theta\varepsilon_{II}+\tilde{\sigma}\varepsilon_{I})
+\Omega\left(\varepsilon_{KL}+\frac{\kappa}{2\tilde{f_R}}(1+\tilde{f_T})\Pi_{KL}\right)
\\\nonumber&+2H_2a_{I}+\frac{1}{B}\left(\frac{H_2C'}{C}+H'_2\right)+\frac{H_1}{rB}
\left(\frac{C_{,\theta}}{C}-\frac{B_{,\theta}}{B}\right)=
-\frac{\kappa}{6}(\Theta+\tilde{\sigma})\left(\tilde{\mu}^{\textrm{eff}}
+(\tilde{P}+\frac{\tilde{\Pi_{II}}}{3})^{\textrm{eff}}\right)
\\\label{73}&-\kappa a_{I}\tilde{q_{I}}^{\textrm{eff}}
-\frac{\kappa}{2\tilde{f_R}}(1+\tilde{f_T})\left[\frac{q_{II}}{Br}+\frac{q_{I}}{B}
\frac{(Br)'}{Br}\right],\\\nonumber
&\frac{1}{3A}\left[-\frac{\kappa}{2\tilde{f_R}}(1+\tilde{f_T})(-\mu+\Pi_{I}
+\Pi_{II})-(\varepsilon_{II}+\varepsilon_{I})\right]^{\dot{}}
+\frac{\kappa}{18\tilde{f_R}}(1+\tilde{f_T})(\Pi_{I}+\Pi_{II})(2\tilde{\sigma}-\Theta)
\\\nonumber&-2(a_IH_2-a_{II}H_1)-\frac{1}{3}(\Theta+\tilde{\sigma})(\varepsilon_{I}+\varepsilon_{II})
+\frac{1}{B}\left(H_2\frac{(Br)'}{Br}+H'_2\right)+\frac{1}{rB}
\left(H_{1,\theta}+\frac{B_{,\theta}}{B}H_1\right)
\\\label{74}&=-\frac{\kappa}{6}(\tilde{\mu}^{\textrm{eff}}
+\tilde{P}^{\textrm{eff}})(\Theta-2\tilde{\sigma})
-\frac{\kappa\tilde{q_{I}}^{\textrm{eff}}}{2B}\frac{C'}{C}
-\frac{\kappa\tilde{q_{II}}^{\textrm{eff}}}{2Br}\frac{C_{,\theta}}{C},\\\nonumber
&\frac{1}{3B}\left(\varepsilon_I+\frac{\kappa}{2\tilde{f_R}}(1+\tilde{f_T})\Pi_I\right)'
+\frac{1}{Br}\left(\frac{\kappa}{2\tilde{f_R}}(1+\tilde{f_T})\Pi_{KL}+\varepsilon_{KL}\right)_{,\theta}
+\frac{1}{3B}\left(\right.\varepsilon_I\frac{\kappa}{2\tilde{f_R}}(1+
\\\nonumber&+\tilde{f_T})\Pi_I\left.\right)
\left(\frac{2C'}{C}+\frac{(Br)'}{Br}\right)
+\frac{1}{3B}\left(\varepsilon_{II}+\frac{\kappa}{2\tilde{f_R}}(1+\tilde{f_T})\Pi_{II}\right)
\left(\frac{C'}{C}-\frac{(Br)'}{Br}\right)
+\frac{1}{Br}
\\\label{75}&\left(\varepsilon_{KL}+\frac{\kappa}{2\tilde{f_R}}(1+\tilde{f_T})\Pi_{KL}\right)
\left(\frac{C_{,\theta}}{C}-\frac{B_{,\theta}}{B}\right)
=\frac{2\kappa}{6B\tilde{f_R}}(1+\tilde{f_T})\mu',\\\nonumber
&\frac{1}{3Br}\left(\varepsilon_{II}+\frac{\kappa}{2\tilde{f_R}}(1+\tilde{f_T})\Pi_{II}\right)_{,\theta}
+\frac{1}{B}\left(\frac{\kappa}{2\tilde{f_R}}(1+\tilde{f_T})\Pi_{KL}+\varepsilon_{KL}\right)'
+\frac{1}{3Br}\left(\right.\varepsilon_I+\frac{\kappa}{2\tilde{f_R}}
\\\nonumber&\times(1+\tilde{f_T})\Pi_I\left.\right)\left(\frac{C_{,\theta}}{C}-\frac{B_{,\theta}}{B}\right)
+\frac{1}{3Br}\left(\varepsilon_{II}+\frac{\kappa}{2\tilde{f_R}}(1+\tilde{f_T})\Pi_{II}\right)
\left(\frac{2C_{,\theta}}{C}+\frac{B_{,\theta}}{B}\right)
+\frac{1}{B}
\\\label{76}&\times\left(\frac{\kappa}{2\tilde{f_R}}(1+\tilde{f_T})\Pi_{KL}+\varepsilon_{KL}\right)
\left(\frac{2C'}{C}+\frac{(Br)'}{Br}\right)
=\frac{2\kappa}{6B\tilde{f_R}}(1+\tilde{f_T})\mu_{,\theta},\\\nonumber
&-\frac{1}{B}\left[H_1\left(\frac{(Br)'}{Br}+\frac{2C'}{C}\right)+H'_1\right]
-\frac{1}{rB}\left[H_{2,\theta}+H_2\left(\frac{B_{,\theta}}{B}+\frac{2C_{,\theta}}{C}\right)\right]
+\frac{\kappa}{2B\tilde{f_R}}(1+\tilde{f_T})
\\\nonumber&\times\left(q_{II}\frac{(Br)'}{Br}
+q'_{II}\right)+\Omega\left[\kappa(\tilde{\mu}^{\textrm{eff}}
+\tilde{P}^{\textrm{eff}})-(\varepsilon_{I}+\varepsilon_{II})+\frac{\kappa}{6\tilde{f_R}}
(1+\tilde{f_T})(\Pi_{I}+\Pi_{II})\right]-\frac{\kappa}{2rB\tilde{f_R}}
\\\label{76}&\times(1+\tilde{f_T})\left(q_{I,\theta}+q_{I}\frac{B_{,\theta}}{B}\right),\\\nonumber
&-\frac{1}{B}\left(\frac{\kappa}{6\tilde{f_R}}(1+\tilde{f_T})\Pi_{KL}\right)'
+\frac{1}{rB}\left(\frac{\kappa}{3\tilde{f_R}}(1+\tilde{f_T})\Pi_{KL}\right)_{,\theta}
-\frac{\varepsilon_I}{3rB}\left(\frac{C_{,\theta}}{C}+\frac{2A_{,\theta}}{A}\right)
-\frac{\varepsilon_{II}}{3rB}
\\\nonumber&\times\left(\frac{2C_{,\theta}}{C}+\frac{A_{,\theta}}{A}\right)
-\frac{\kappa}{B\tilde{f_R}}(1+\tilde{f_T})\Pi_{KL}\frac{(Br)'}{Br}-\frac{\varepsilon_{KL}}{B}
\left(\frac{C'}{C}-\frac{A'}{A}\right)+\frac{\kappa}{6rB\tilde{f_R}}(1+\tilde{f_T})(\Pi_{I}
\\\label{77}&-\Pi_{II})\frac{B_{,\theta}}{B}+\frac{1}{A}\dot{H_1}
=-\frac{2\kappa}{6B\tilde{f_R}}(1+\tilde{f_T})\mu_{,\theta},\\\nonumber
&-\frac{1}{B}\left(\frac{\kappa}{6\tilde{f_R}}(1+\tilde{f_T})\Pi_{II}
-(\varepsilon_{I}-\varepsilon_{II})\right)'+\frac{\kappa}{2rB\tilde{f_R}}(1+\tilde{f_T})
\left(\Pi_{KL}\frac{2B_{,\theta}}{B}+\Pi_{KL,\theta}\right)+\frac{1}{A}\dot{H_2}
\\\nonumber&+\frac{\varepsilon_{I}}{3B}\left(\frac{2C'}{C}+\frac{A'}{A}\right)
+\frac{\varepsilon_{II}}{3B}\left(\frac{C'}{C}+\frac{2A'}{A}\right)
+\frac{\kappa}{6B\tilde{f_R}}(1+\tilde{f_T})(\Pi_{I}
-\Pi_{II})\frac{(Br)'}{Br}+\frac{\varepsilon_{KL}}{rB}\left(\right.\frac{C_{,\theta}}{C}
\\\label{78}&\left.-\frac{A_{,\theta}}{A}\right)
=\frac{\kappa}{6B\tilde{f_R}}(1+\tilde{f_T})\mu',
\end{align}
where $\tilde{q_{II}}^{\textrm{eff}}$ shows the quasi static-evolution of effective
component of heat-flux $(q_{II}^{\textrm{eff}})$. So,
one can easily computed the quasi static-evolution of effective components
of relativistic-fluid by using the conditions of the QSA as
defined in above section.

\section{Appendix B}

The expression of effective fluid contents for our relativistic system are
\begin{align}\nonumber
&\mu^{\textrm{eff}}=\frac{1}{f_R}\left[\right.\tilde{\mu}-\frac{1}{2}(f-Rf_R)+\chi_0\left.\right],\quad
q^{\textrm{eff}}_I=\frac{1}{f_R}\left[\right.q_I(1+f_T)-\frac{1}{AB}\chi_1\left.\right],\\\nonumber
&q^{\textrm{eff}}_{II}=\frac{1}{f_R}q_{II}(1+f_T)-\frac{1}{f_R\sqrt{r^2A^2B^2+G^2}}\left[\right.
G\chi_0+\chi_2\left.\right],\\\nonumber
&\Pi^{\textrm{eff}}_{KL}=\frac{1}{f_R}(1+f_T)\Pi_{KL}+\frac{1}{f_R\sqrt{r^2A^2B^2+G^2}}\left[\right.
G(q_I
f_T-\frac{1}{AB}\chi_1)+\frac{A}{B}\chi_3\left.\right],\\\nonumber
&(P+\frac{\Pi_I}{3})^{\textrm{eff}}=\frac{1}{f_R}(1+f_T)(P+\frac{\Pi_I}{3})+\frac{1}{f_R}\left[\right.
\tilde{\mu}
f_T+\frac{1}{2}(f-Rf_R)+\frac{1}{B^2}\chi_4\left.\right],\\\nonumber
&(P+\frac{\Pi_{II}}{3})^{\textrm{eff}}=\frac{1}{f_R}(1+f_T)(P+\frac{\Pi_{II}}{3})
+\frac{1}{f_R\sqrt{r^2A^2B^2+G^2}}\left[\right.\{\frac{1}{2}(f-Rf_R)
(G-rB^2)-G\chi_0\}
\\\nonumber&+2\{\sqrt{r^2A^2B^2+G^2}q_{II}f_T
+G(\chi_0+\frac{\chi_2}{r})\}+rB^2(\tilde{\mu}
f_T+\frac{\chi_5}{C^2})\left.\right].
\end{align}
The values of $\chi_i's$ appearing in the expression of effective fluid contents and in
Eqs.\eqref{41} are presented in Appendix in\cite{yousaf2021axially}.

The extra terms $\psi_i's$ appearing in
Eqs.\eqref{22}, \eqref{25}, \eqref{26} and \eqref{28} are
\begin{align}\nonumber
&\psi_1=\frac{\kappa}{8f_R}\epsilon_{\lambda}^{\epsilon \gamma}
\left[\right.(\nabla^{\pi}\nabla_{\epsilon}f_R) \epsilon_{\omega\pi
\gamma}-(\nabla^{\pi}\nabla_{\gamma}f_R)\epsilon_{\omega\pi\epsilon}
-(\nabla^{\alpha}\nabla_{\epsilon}f_R)\epsilon_{\alpha\omega\gamma}
+(\nabla^{\alpha}\nabla_{\gamma}f_R)\epsilon_{\alpha\omega\epsilon}\left.\right],\\\nonumber
&\psi_2=\nabla_{\lambda}\nabla_{\omega}f_R-\frac{3}{2}
U_{\lambda}U_{\omega}(f-Rf_R)-\nabla_{\lambda}\nabla_{\gamma}f_RU^{\gamma}U_{\omega}
+2U_{\lambda}U_{\omega}\Box
f_R-(\nabla^{\alpha}\nabla_{\omega}f_R)U_{\lambda}U_{\alpha}
+g_{\lambda\omega}
\\\nonumber&\times(\nabla^{\alpha}\nabla_{\gamma}f_R)U_{\alpha}U^{\gamma},\\\nonumber
&\psi_3=\nabla^{\lambda}\nabla_{\lambda}f_R+\frac{3}{2}(f-Rf_R)
-(\nabla^{\lambda}\nabla_{\gamma}f_R)U^{\gamma}U_{\lambda}
-2\Box f_R-(\nabla^{\alpha}\nabla_{\lambda}f_R)U^{\lambda}U_{\alpha}
+4(\nabla^{\alpha}\nabla_{\gamma}f_R)U_{\alpha}U^{\gamma},\\\nonumber
&\psi_4=(\nabla^{\alpha}\nabla_{\gamma}f_R)U^{\gamma}\epsilon_{\lambda
\alpha\omega}.
\end{align}

\end{document}